\documentclass[twocolumn]{bmcart}

\usepackage[utf8]{inputenc} 

\def\includegraphics{}

\startlocaldefs
\def\var{\mbox{Var}}

\def\trace{\mbox{ trace}}

\def\etal{{\em et al.}}
\def\vs{\mbox{vs.}}
\endlocaldefs

\begin{document}
\begin{frontmatter}

\begin{fmbox}
\dochead{Research}

\title{Analysing multiple types of molecular profiles simultaneously: connecting the needles in the haystack}
 

\author[
   addressref={aff1},                   
   corref={aff1},                       
   email={r.menezes@vumc.nl}   
]{\inits{RX}\fnm{Ren\'{e}e X} \snm{Menezes}}
\author[
   addressref={aff2},
   email={mohammadileila@yahoo.com}
]{\inits{L}\fnm{Leila} \snm{Mohammadi}}
\author[
   addressref={aff3},
   email={jelle.goeman@radboudumc.nl}
]{\inits{JJ}\fnm{Jelle J} \snm{Goeman}}
\author[
   addressref={aff4,aff5},
   email={j.m.boer@erasmusmc.nl}
]{\inits{JM}\fnm{Judith M} \snm{Boer}}


\address[id=aff1]{
  \orgname{Department of Epidemiology and Biostatistics, VU University Medical Center}, %
  \street{De Boelelaan 1089a},                     %
  \postcode{1081 HV}                                
  \city{Amsterdam},                              
  \cny{The Netherlands}                                    
}
\address[id=aff2]{%
  \orgname{Department of Statistics},
  \city{Shiraz},
  \cny{Iran}
}
\address[id=aff3]{%
  \orgname{Biostatistics, Department for Health Evidence, Radboud University Medical Center},
  \city{Nijmegen},
  \cny{The Netherlands}
}
\address[id=aff4]{%
  \orgname{Department of Pediatric Oncology and Hematology, Erasmus MC-Sophia Children's Hospital},
  \city{Rotterdam},
  \cny{The Netherlands}
}
\address[id=aff5]{%
  \orgname{Netherlands Bioinformatics Centre},
  \city{Nijmegen},
  \cny{The Netherlands}
}


\begin{abstractbox}

\begin{abstract} 
\parttitle{Background} 
It has been shown that a random-effects framework
can be used to test the association between a gene's expression level and the number of DNA copies of a set of genes.  
This gene-set modelling framework was later applied to find associations between mRNA expression and microRNA expression, by defining the gene sets using target prediction information. 

\parttitle{Methods and Results}
Here, we extend the model introduced by Menezes \etal(2009) 
to consider the effect of not just copy number, but also of other molecular profiles such as 
 methylation changes and loss-of-heterozigosity (LOH), on gene expression levels. We will consider again sets of measurements, to improve robustness of results and increase the power to find associations. Our approach can be used genome-wide to find associations, yields a test to help separate true associations from noise and can include confounders. 

We apply our method to colon and to breast cancer samples, for which genome-wide copy number, methylation and gene expression profiles are available. Our findings include interesting gene expression-regulating mechanisms, which may involve only one of copy number or methylation, or both for the same samples. We even are able to find effects due to different molecular mechanisms in different samples. 

\parttitle{Conclusions}
Our method can equally well be applied to cases where other types of molecular (high-dimensional) data are collected, such as LOH, SNP genotype and microRNA expression data. Computationally efficient, it represents a flexible and powerful tool to study associations between high-dimensional datasets. The method is freely available via the SIM BioConductor package.
\end{abstract}



\end{abstractbox}
\end{fmbox}

\end{frontmatter}



\section*{Background}
The wealth of omics data being currently produced raises the need for efficient and flexible models to analyse these data. One typical objective is to understand which molecular 
changes affect gene expression levels or, if available, protein expression levels. Molecular profiles reliably measured currently include DNA methylation and copy number, sequence information including SNP and loss-of-heterozigosity (LOH) information, as well as microRNA expression levels. All these are known to be involved in gene expression regulation.

Many methods have so far been proposed for finding associations between two omics data sets (for an overview see \cite{huang2012}). These enable us to study for example which copy number changes affect gene expression levels, or alternatively which methylation changes affect gene expression, and to which extent they do so. Such findings can for example shed light onto oncogenic mechanisms.   One such method has been proposed by Menezes {\sl et al.}\cite{menezes2009}, whereby a statistical  test is used to for example assess the association between gene expression and the copy number of a genomic region around it. In that work, the authors showed the power gain to find associations yielded by considering sets of measurements, rather than considering only associations between pairs of features, as done elsewhere\cite{stranger2007}.

Methods have also been proposed for the joint analysis of microRNA expression and mRNA expression profiles. In this context, many microRNAs can target the same mRNA, and one microRNA can target multiple mRNAs. So it makes sense to consider target prediction information while looking for associations between mRNA and microRNA expression levels. van Iterson {\sl et al.}\cite{iterson2013} use the method proposed by Menezes {\sl et al.}\cite{menezes2009}, with gene sets now defined by various target prediction tools. It is then shown that the method has power to find associations that are sucessfully validated, in spite of limited sample size. They also show that predicted associations using sets of features lead to more robust, and thus more reliable, results, in this case with a higher validation rate, than associations predicted using pairs of features (i.e. one miRNA and one mRNA at a time). This is in agreement of findings from Menezes\etal\cite{menezes2009}. The extra robustness is brought in by the fact that the impact on small effects observed for specific genes must be observed for at least a considerable subset of genes, before they become significant.

With multiple types of molecular profiles measured, it makes sense to consider methods for analysing all of them together. 
Only a few of the integrated analysis methods so far proposed have been extended to handling more than two data sets. Methods proposed by Waaijenborg \etal\cite{waaijenborg2009} and by Witten \etal\cite{witten2009} use a sparse canonical correlation framework. As such, they are of an exploratory nature, aiming at finding sets of covariates from the various data sets which are most correlated. 

Here we extend the integrated analysis method proposed by Menezes \etal\cite{menezes2009} to handle multiple high-dimensional data sets. The aim is to test for association between one type of molecular profile (mRNA, say) and other types (copy number and methylation, say), the latter represented by sets of probes, rather than individual ones. Under the null hypothesis, no association exists between mRNA and either copy number or methylation, in our example. The use of sets makes for a robust and powerful method: robust, because noise originating from individual probes is ignored; and powerful, because subtle associations found between mRNA and multiple methylation probes are detectable as the probability of seeing many of these small associations together by chance will be considered small, which would be ignored if they were considered separately. Given the high-dimensionality of the problem, the fact that our method evaluates statistical significance is crucial to help separate noise from true associations. Moreover, since our method uses a regression framework, it can take confounders into account. As the original method, it is thus a flexible, powerful and efficient method to analyse jointly many omics data sets. 

This paper is organized as follows. In the Methods section we present the statistical test for associations between multiple gene sets and a response. In the Results section we illustrate the workings of our method under various types of associations between data sets with a simulation study. We also apply our method to  sets of TCGA samples: 125 of colon cancer  and 173 of breast cancer, for which copy number, methylation and gene expression profiles are available.

\section*{Methods}

\subsection*{The integrated analysis model}


Menezes \etal\cite{menezes2009} have proposed using score tests to find associations between a response (say, the expression levels of a gene) and a set of covariates (say, the genomic copy number measured at multiple loci on the same chromosome arm as the gene). Let us represent by $Y_{ni}$ the 
 expression for gene $i$, and by $X_{nj}$ the genomic copy number for gene $j$, where $i=1,\dots,I$ and $j=1,\dots,J$ represent the sets of probes used, and $n=1,\dots,N$ indexes the sample.   Note that it is assumed that measurement sets $\{Y_i\},\{X_j\}$ are available per sample, although it is not necessary that both sets of probes correspond to the same loci. Then we write, for any given $i$,

\begin{equation}
E\left(Y_{ni}\right) = h\left( \alpha + \sum_{j=1}^J \beta_j X_{nj} \right),\ n=1,\dots,N,
\label{sim.1gs}
\end{equation}
where $h(.)$ is a given inverse link function and $\{\beta_j\}\sim{\cal N}(0,\tau)$, making model (\ref{sim.1gs}) a random-effects model. From now on, we consider observations ${Y_i}$ for a single gene indexed by $i$, but we omit the index $i$ for clarity. 

The approach proposed by Menezes \etal\cite{menezes2009} focuses not on fitting model (\ref{sim.1gs}) directly, but on testing whether or not the association between $Y$ and $\{X_j\}$ is statistically significant, for each gene expression probe $Y$. This is done by making use of the global test \cite{goeman2004}. In this framework, under the null hypothesis that $Y$ is not associated with the set $\{X_j\}$, we have $\var(\beta)=\tau=0$. On the other hand, when $Y$ displays association with variables in $\{X_j\}$, then some of the $\beta_j$ will be non-zero, and thus $\var(\beta)>0$.  Specifically, the global test is a score test for the hypotheses 

$$
H_0: \var(\beta) = 0 \quad \vs\quad H_a: \var(\beta) \neq 0,
$$
first proposed by \cite{lecessie1995} and later applied to the context of testing association between a molecular profile and a clinical variable by Goeman \etal\cite{goeman2004}. Goeman also further extended the test to generalized linear models and Cox regression models \cite{goeman2005}, and showed that it is the locally most powerful test against a set of alternatives often of interest \cite{goeman2006}. Let us define $r_n=\sum_j \beta_j X_{nj}$, the part of the linear predictor that depends on the data, $r=(r_1,r_2,\dots,r_N)^t$, and $X$ an $N\times J$ matrix containing all observations for the covariates. Then a statistic to test the hypotheses above was proposed by Goeman \etal\cite{goeman2004} as

\begin{equation}
 Q(X) \propto \frac{(Y-\mu)^t X X^t (Y-\mu)}{(Y-\mu)^t (Y-\mu)},
\label{qx.1gs}
\end{equation}
or in its standardized form

\begin{equation}
T = \frac{Q(X)- E\left[Q(X)\right]}{\sqrt{\var\left[Q(X)\right]}}.
\label{test.1gs}
\end{equation}
Goeman \etal\cite{goeman2004} have also shown that $E(Q)=\trace(X X^t)/J$ and $\var(Q) = 2\trace[(X X^t)^2]/J^2$. As $Q$ is a (scaled) quadratic form, its distribution is asymptotically equivalent to a weighted sum of 1 d.f. chi-square distributions (see \cite{goeman2011}) and, for some special cases such as the linear model, numerical integration can be used to find its cumulative distribution function in finite-sample problems \cite{goeman2006}. Note that the numerator of the test statistic $Q(X)$ is a quadratic form involving the covariances between the random effects $\{r_n\}$, given by $X X^t$ up to a constant. 

Applying this test generates one p-value per gene expression variable $Y$. For a set of gene expression probes $\{Y_i,\ i=1,\dots,I\}$ a list of p-values is obtained, where $i$ indicates different probes, loci or genes. Multiple-testing correction must be applied before drawing conclusions. Individual contributions of each copy number variable $X_j$ on the dependent variable $Y_i$ can be computed, generating an overview of association patterns between copy number and gene expression profiles.

\subsection*{Extension to two sets of covariates}

Assume now that a second set of covariates $\{Z_k\}$ is observed, and there is interest in studying the association between both covariate sets and the response $Y_i$. Now model (\ref{sim.1gs}) becomes

\begin{equation}
E\left(Y_{ni}\right) = \alpha + \sum_{j=1}^J \beta_j X_{nj}+ \sum_{k=1}^K \gamma_k Z_{nk},\ n=1,\dots,N,
\label{sim.2gs}
\end{equation}
where $\{\beta_j\}\sim{\cal N}(0,\tau)$ and $\{\gamma_k\}\sim{\cal N}(0,\theta)$, so that model (\ref{sim.2gs}) still is a random-effects model, and where we have assumed for simplicity that $h(x)=x$. Similarly to the single covariate-set case, under the null hypothesis no association exists between either $Y$ and $\{X_j\}$, or $Y$ and $\{Z_k\}$. In this case, obviously the variances $\tau,\theta$ of the random effects $\{\beta_j\},\{\gamma_k\}$ in (\ref{sim.2gs}) must be zero. On the other hand, under the alternative hypothesis, $Y$ displays association with either $\{X_j\}$ or $\{Z_k\}$, meaning that either one of the two random effect variances $\tau,\theta$ must be non-zero. Thus, to test for association between each $Y_i$ and the sets of variables $\{X_j\},\{Z_k\}$, it is of interest to test the hypotheses 

$$
H_0: \var(\beta) = \var(\gamma) = 0 \quad \vs\quad H_a: \var(\beta) \neq 0\ \mbox{or}\  \var(\gamma) \neq 0
$$
By proceeding in a similar way as in le Cessie \& van Houwelingen\cite{lecessie1995} and in Goeman \etal\cite{goeman2004} for the one covariate set case, we can write the test statistic for these hypotheses as 

\begin{equation}
T_{XZ}^2 = \frac{T^2_X + T^2_Z - 2 \rho_{XZ} T_X T_Z}{1-\rho_{XZ}},
\label{test.2gs}
\end{equation} 
where $T_X,T_Z$ are the standardized test statistics computed with only one covariate set in the model defined by (\ref{test.1gs}), and $\rho_{XZ}$ is the correlation between $T^2_X$ and $T^2_Z$. A detailed derivation is given in section 2 of the appendix. Note that if in (\ref{test.2gs}) we take $\rho_{XZ}\equiv 0$, we obtain the simpler expression

\begin{equation}
T_{XZ}^2 = T^2_X + T^2_Z.
\label{tx.2gs.nocor}
\end{equation} 
For simplicity, we may ignore the standardization and the squaring, then write the unscaled test statistic for two gene sets as

\begin{equation}
Q(X,Z) \equiv Q(X) + Q(Z).
\label{qx.2gs}
\end{equation} 
This makes sense. Indeed, from (\ref{qx.1gs}), 

$$
Q(X) + Q(Z) = \frac{(Y-\mu)^t \left[  X X^t  +  Z Z^t  \right] (Y-\mu)}{(Y-\mu)^t (Y-\mu)},
$$
where  $X X^t + Z Z^t$ is the matrix in the quadratic form that would have been obtained if our model had a single set of covariates given by the merged set $\{X,Z\}$, and with effect modelled by a single vector of random effects. In such a case, the design matrix would have been obtained by binding the columns of $X$ and $Z$ together and, thus, the unscaled score test statistic would be given as above. 

Thus, the standardized test statistic for two covariate sets is equal to the sum of the individual test statistics per covariate set, if the correlation $\rho_{XZ}$ between the test statistics can be ignored. Similarly for the unscaled and unsquared versions of the score test statistics, the test statistic for two gene sets $X$ and $Z$ is equivalent to the one obtained for a single gene set formed by $\{X,Z\}$.

In practice, some sort of centering and scaling of $Q(X),Q(Z)$ may be used, especially when covariates take values in different ranges, are of very different sizes and/or display different variances. In such cases, centering and scaling can ensure the separate sets are given the correct weights in the combined test statistic $Q(X,Z)$. This can be done by using the scaled individual test statistics as in (\ref{tx.2gs.nocor}). An alternative is given in Section 5 of Additional File 1.

\subsection*{Extension to more than two gene sets}

A closed analytical form for the score test statistic can be obtained for any given number of gene sets $M$, as shown in section 3 of the appendix. Here we note that the score test statistic typically is a function of the individual test statistics and their pairwise correlations. If the pairwise correlations are ignored, the squared test statistic for $M$ gene sets can be written as the sum of the squared and standardized test statistics for the individual gene sets, generalizing the result obtained for $M=2$. In particular, we note that if unsquared and non-standardized versions of the test statistics are used, it is easy to see that the test statistic for any $M$ number of gene sets is the same one as generated by a model with a single gene set, formed by merging all gene sets together.

\subsection*{The role of the correlation}

The correlation $\rho_{XZ}$ between test statistics for single gene sets $T_X,T_Z$ is an integral part of the score test statistics for multiple gene sets, as seen above. Although such correlations can be estimated, for example via re-sampling, this makes computations considerably more complex. It turns out that correlations between individual test statistics can be ignored with little loss of power. Here we give a heuristic argument for this, and in the Results section we will confirm this in a simulation study.

We have seen in expression (\ref{qx.2gs}) that, if we consider the unscaled and linear form of test statistics, the score test statistic for two gene sets $Q(X,Z)$ equals the test statistic for a single gene set, formed by the union of all gene sets into a single one, or $Q(W)$ where $W$ is a matrix formed by the columns of $X,Z$ bound together. As such, the test statistic already considers not only correlation between the covariates in the set and the dependent variable $Y_i$, but also pairwise correlations between the covariates in the set, as shown by Goeman \etal\cite{goeman2006}, section 7. Thus, correlations between the covariates, and between the covariates and the dependent variable, are already considered and catered for by the score test statistic. Therefore, the explicit inclusion of the correlations between test statistics for individual sets is not essential, as these correlations are already taken into account. 

For this reason, and due to its simplicity, we suggest ignoring the correlation when computing the joint test statistic. This means using either the unscaled test statistic (\ref{qx.2gs}) or its scaled version (\ref{tx.2gs.nocor}). We will verify in simulation studies that the test statistics with and without the correlation yield similar power, in situations of practical interest.

\subsection*{Test statistic null distribution} 

For testing, the distribution of $T_{XZ}^2$ under the null hypothesis of no association between $Y_i$ and $X,Z$ is needed. Here we will consider the expression (\ref{qx.2gs}) for the combined test statistic. For the single-set testing, Goeman \etal\cite{goeman2011} obtains an expression for the asymptotic distribution of $Q(X)$ under a generalized linear model, which is the exact finite sample distribution under the linear model, as used here. This (asymptotic) null distribution can be written as a ratio of weighted sums of $\chi^2_1$ random variables\cite{goeman2011}. 

Our test statistic (\ref{qx.2gs}) for two covariate sets $X,Z$ can be seen as a test statistic for a single covariate set resulting of the union of the two original sets. This means that the distributions derived in Goeman \etal\cite{goeman2011} can be used for (\ref{qx.2gs}).    

Note that, in case of applying the test to many separate responses $\{Y_i\}$, such as expression levels of many different genes, the resulting computational burden of numerically estimating the distribution per response may be superior to computing p-values via permutation. In the cancer examples, we use permutations to compute p-values for $Q(X,Z)$ for computational ease, and use the sum of test statistics (\ref{qx.2gs}).

\subsection*{Software}

Methods presented in this work are implemented in the Bioconductor package SIM, currently for a single covariate set, and in the short-term for multiple covariate sets. All computations described and applied in this work used R from at least version 3.0.1 (see \cite{refR2015} for a recent reference).


\section*{Results}

\subsection*{Simulation study 1}


We run a simulation study to evaluate the power of the proposed test statistics under various types of effects. A detailed description of the study setup is given in Section 1 of the Additional File 2. For completeneness, here follows a brief description. The data here is assumed to consist of two explanatory sets of covariates, $\{X_j,\ j=1,\dots,J\}$ and $\{Z_k,\ k=1,\dots,K\}$, and a set of dependent variables $\{Y_i,\ i=1,\dots,I\}$. For simplicity we assume that $I=J=K$. 

We consider four independent data sets, each involving one set of variables $\{Y_i,X_j,Z_k\}$,  with $i,j,k=1,\dots,1000$. Each data set can be seen as a (genomic) region, here assumed to involve one association type between the covariate sets and the dependent variable $Y_i$ for $i=1,\dots,500$, and no association for the remaining probes. The associations considered are: region I, where $Y_i$ is associated with $\{X_j\}$ only, which we will refer to as ``x only''; region II, ``additive'', where both covariate sets affect outcome linearly; region III, ``multiplicative'', where both covariate sets affect outcome linearly as well as multiplicatively; and region IV, ``split-samples'' or ``complementary'',  where $\{Y_i\}$ depends upon $\{X_j\}$ for half of the samples, and for the other half $\{Y_i\}$ depends upon $\{Z_k\}$.   For each data set, three sample sizes are considered: 50, 100 and 200 samples.

Within each region, we test for association between each $Y_i$ and the covariate sets $\{X_j\},\{Z_k\}$ using both the test statistics for two gene sets (\ref{test.2gs}), where correlation is estimated via permutation, as well as its simplified expression by ignoring the correlation (\ref{tx.2gs.nocor}). Note that these are the scaled versions of the test statistics, as it is in the expression for $T^2_{XZ}$ that the correlation $\rho_{XZ}$ between $T_X^2$ and $T_Z^2$ appears.
Here p-values are estimated by comparing the observed test statistic to values obtained after permuting the dependent variable samples 1000 times, and re-computing the test statistics. For the aims of this particular study, which is to compare ROC curves, no multiple testing is necessary.


ROC curves were made to evaluate the power of finding the four effect types (figure 1 for $N=100$ - see supplementary figures 1 and 2 in Additional File 2 for results corresponding to sample sizes 50 and 200). In all cases, the power yielded with the test statistic with correlation (\ref{test.2gs}), indicated in purple, is virtually the same and the one with the test statistic that ignores correlation (\ref{tx.2gs.nocor}), indicated in pink. This confirms our arguments, given in the Methods section, that correlation between test individual test statistics need not be taken into account as it is already intrinsically included. Test statistics used displayed acceptable power to find effects, with the most difficult effect to find being the complementary. This is related to the nature of the effect, where one covariate set explains the dependent variable for a few samples, and the second covariate set explains it for another set of samples. Even in this case, associations can be found if the sample size is relatively large enough. In our setup, the sample size of 50 yields power just above random to find effects, with a clear improvement with $N\geq 100$.

\subsection*{Simulation study 2}


We have argued that correlation between single-set test statistics ($\rho_{XZ}$ in equation \ref{test.2gs}) needs not be taken into account in the two-set test statistic expression, and we have seen in the above simulation study that, indeed, test statistics with and without explicitly including this correlation yield virtually the same power to find a variety of effects. However, our simulation study involved covariate sets $X,Z$ simulated independently, so that no significant correlation between these sets, and ultimately between their corresponding single-set test statistics, would have been expected. So the question remains of whether correlation would have an impact on power to find effects, if covariate sets were correlated. 

To answer this question, we perform a second simulation study where the covariate sets are correlated. Here we use the same setup as in the first simulation study presented in the previous section, with the same regions displaying the same effects. The only difference is that here $Z_j=X_j+U_j$, where the $\{U_j\}$ are independently and identically distributed normal variables with mean 0 and standard deviation 1. This leads to a relatively high positive correlation between $\{X_j\}$ and $\{Z_k\}$, as we can see from figure 2 for $N=100$. To find associations, we use the joint test statistic including the estimated correlation (\ref{test.2gs}), as well as the simplified test statistic given by the sum of the non-scaled single-set  test statistics (\ref{qx.2gs}). 


Our simulation scheme produces highly correlated data sets. The relatively high correlation between samples (blue curve in supplementary figure 3) becomes diluted when test statistics for individual sets are computed (red curve in supplementary figure 3). This is due to the fact that individual-set test statistics measure association with the dependent variables $\{Y_i\}$. In cases where $\{X_j\}$ and $\{Z_k\}$ display similar association with $\{Y_i\}$, such as in the additive region, correlation between $Q(X)$ and $Q(Z)$ will be more similar to that between the covariate sets. However, in cases where the sets $\{X_j\}$ and $\{Z_k\}$ do not display similar association with the dependent variables, correlation between $Q(X)$ and $Q(Z)$ will be almost completely dampened, compared to that between covariate sets directly. Thus, dampening of the induced correlation depends on the region/type of association considered. Nevertheless, some correlation between individual test statistics remains. 

In spite of the correlation between single-set test statistics, ROC curves produced with the test statistics including the estimated correlation (\ref{test.2gs}),  as well as with the test statistic ignoring the correlation ({tx.2gs.nocor}), are virtually identical under additive or split-samples/complementary effects (figure 2 for $N=100$, see supplementary figures 4 and 5 in Additional File 2 for $N=50, 200$ respectively). This can be interpreted in two ways. Firstly, as already discussed in the Methods section, the sum of single-set test statistics $Q(X), Q(Z)$ is equivalent to the test statistic for the merged set $Q(X,Z)$. As such, correlation between covariates, as well as between covariates and response $Y_i$, are implicitly taken into account by the test statistic. This means that there is no added information in including the computed correlations explicitly in the test statistic. Secondly, this result suggests that the test statistics' correlations induce similar shifts in the null and alternative distributions of the joint test statistics, in such a way that power is ultimately not affected. Thus, by disregarding the correlation there is no loss of power to find associations between gene sets and responses. These conclusions can be drawn for all sample sizes used here ($N=50, 100, 200$).


Still from figure 2, we notice that, if covariate sets affect the response in a multiplicative way, then there is a small gain on overall power by taking the correlation into account. However, the difference is negligible for $p\leq 0.03$, where interest mostly lies in practice. Indeed,  since p-values here  have not been corrected for multiple testing, significant associations will be typically found to have uncorrected p-values in this range. This conclusion can be extended for the other sample sizes considered here ($N=50, 200$, see supplementary figures 4 and 5 in Additional File 2).

On the other hand, when a single covariate set is associated with the response $\{Y_i\}$ and the covariate sets are correlated with each other, then to take the correlation into account leads to a power loss. This is likely due to the fact that, although the covariate sets are correlated and, thus, their individual test statistics $Q(X)$ and $Q(Z)$ are also correlated, only $\{X_j\}$ influences the responses $\{Y_i\}$. So, in this case by taking the correlation explicitly into account, noise is introduced in the test statistic, leading to a power loss. This power loss due to using $Q(X,Z)$ is larger than this test statistics' power gain under a multiplicative effect, and it is already seen for p-values very close to zero. 

The results of this simulation study suggest that the simplified test statistic, that does not explicitly include the correlation between single-set test statistics, yields at least as high power as the test statistic that does explicitly include the correlation, for the effect types considered, except for the multiplicative effect, where it leads to a slight power loss. Here we point out that the power loss is noticeable for p-values greater than a threshold larger than zero, although it is not possible to indicate in practice where this threshold would be.

\subsection*{Colon and breast cancer datasets}

\subsubsection*{Data and definitions}

In our simulation studies we have shown that the joint test statistic $Q(X,Z)$ can test for, and find, different types of associations between covariate sets and responses. Here we will see that its relationship with the individual test statistics $Q(X),Q(Z)$ can help elucidate the relative effects of molecular mechanisms under study on the response. To illustrate this, we will consider 
colon and breast cancer data sets extracted from The Cancer Genome Atlas (TCGA). We downloaded 125 colon and 173 breast cancer samples, which had been profiled for DNA copy number (CN), methylation (ME) and gene expression. DNA copy number data was derived from the Affymetrix SNP 6.0 microarray and had been segmented. We de-segmented it to produce individual intensities measured over $3\times 10^4$ equally-spaced positions on the genome. The level-2 DNA-methylation profiles downloaded had been produced by the Infinium Human Methylation 27K array, which produces per probe a {\sl m} and a {\sl u} signal, representing methylated and unmethylated signals respectively. As measurement, we used the logit of the ratio $m/(m+u)$, to transform the ratios into real-valued variables as well as correct them for DNA dosage. The level-2 gene-expression profiles downloaded correspond to lowess-normalized log-ratios obtained using Agilent 450A arrays, involving 62335 probes with known genome location. 

Here we will consider gene expression as the response variable, and we will study how DNA copy number and methylation affect expression levels, both individually as well as jointly. Specifically, we consider a model such as (\ref{sim.2gs}), where $Y_i$ represents the $i$th gene expression probe, $\{X_j\}$ represents the set of copy number probes that are within 1Mb in either direction of the transcription start site of the gene, and $\{Z_k\}$ represents the set of methylation probes that are within 50Kb of the gene's transciption start site in either direction. These window sizes are arbitrary and  reflect current knowledge on the distance of {\sl cis\/}--regulatory effects of copy number (\cite{stranger2007},\cite{blackburn2015})  and methylation on gene expression \cite{bell2011}.

We computed p-values for $Q(X,Z)$, which we will refer to as the joint test, as well as for  the individual test statistics $Q(X),Q(Z)$, which will be referred to as the CN test and the ME test, for association between DNA copy number or  methylation with the gene expression.  
In all comparisons, we use p-values not corrected for multiple testing, as the comparisons between different test results can be more reliably done in this way (in practice, multiple testing should always be used).  Associations for which $p\leq 0.001$ were selected. In all cases, empirical p-values distributions were verified to be enriched with small p-values, so that in none of the cases does the set of mRNA probes selected have small p-values entirely due to chance. Various ratios of the number of selected tests will be computed, following definitions in supplementary table 2 (Additional File 2).

Tables of all mRNA probes tested, with p-values computed by joint and individual tests, can be found in Additional File 3 for the colon cancer data, and Additional File 4 for the breast cancer data.

\subsubsection*{Individual and joint copy number and methylation effects}

Individually, copy number changes explain a larger portion of the gene expression variance than methylation, notably for the colon  cancer data set (see supplementary table 3  in Additional File 2). Interestingly, in the colon cancer data we find twice as many mRNA expression probes  selected as associated with copy number (16\% of the total) than in the breast cancer data (8\%),  despite the latter involving almost 50\% more samples than the former (173 and 125, respectively). So, copy number changes regulate the expression of a larger number of genes for colon cancer, compared with breast cancer. The proportion of mRNA expression probes selected as being regulated by methylation changes is comparable in these two data sets (5.4 and 5.2\% for colon and breast, respectively).

Results separately per chromosome arm mostly reflect the stronger copy number effect in colon cancer compared with breast cancer, although the relative difference between colon and breast cancer here varies depending on the chromosome arm. Methylation effects are, on the other hand, observed in very similar proportions (figure 3 and supplementary tables 4 and 5). Examples of this are found with 1q and 17q. 
However, two chromosome arms are more extreme: 20q and, to a lesser extent 13q, are outliers in the colon cancer data set, with a larger proportion of selected associations with either copy number or methylation than the remaining arms. 

The relationship between the  proportions of selected joint tests for colon and breast cancer (supplementary figure 6, Additional File 2) reflect mostly the relationship found with the CN test (figure 3, left graph); this is due to the copy number effect largely influencing the joint effect.

We can better understand these differences if we now look at the relationship between tests gene-wise, i.e. considering test results per gene. For colon cancer, the joint test yields virtually the same results as the CN test, for 20q and 13q in colon cancer (see top-left hand-side graphs of supplementary figures 7 and 8). In addition, more than 76\% of all mRNA probes selected by ME tests are also selected by CN tests, so that copy number explains  on its own most of gene expression variability (supplementary figure 9). In contrast, for breast cancer the joint test selects mRNA probes that are not selected by the CN test on 20q (supplementary figure 7, top-right), and only between 26 and 37\% of the mRNA probes selected by ME tests are also selected by CN tests (supplementary figure 9 and supplementary table 6). The fact that, in colon cancer, methylation effects mostly overlap with copy number effects cannot be due to measurement artefact, as the methylation measurements are corrected for total DNA. This suggests that copy number drives the associations and, as both 13q and 20q often display copy gains in colon cancer (which is the case in this data), it follows that those gains are likely to be important for oncogenesis.


In contrast, 5q, 16p and 16q display a relatively high proportion of associations with copy number, but not with methylation, for both cancer types (figure 3). Here, the joint test selects additional mRNA probes, not selected by the individual test statistics (see supplementary figure 10 for 5q -- results for 16p and 16q are similar and not shown). In addition, between 18 and 35\% of the mRNA probes selected by ME tests are also selected by the CN test in colon cancer, whereas in breast cancer these ratios lie between 52 and 64\% (supplementary figure 9). This makes results for 13q and 20q in colon cancer more remarkable: they yield the highest percentages of selected CN tests and ME tests, and the largest overlap of ME tests with CN tests.

Comparisons between genome-wide or chromosome arm-wide proportions help picking up trends that differentiate these two cancer types, although they obviously ignore region-specific effects: 17q has commonly amplified regions in breast cancer with an impact on gene expression \cite{menezes2009}, which is not the case for colon cancer.

\subsubsection*{Overlap between individual and joint tests}

Since joint test statistics are the sum of individual ones, results from joint and individual tests naturally may display some association. The extent of this association varies, as it indicates how much of the total gene expression variability is explained by each individual effect.
In order to study this, we will consider the proportion of tests selected both individually as well as with the joint test, compared to the total of selected joint tests. These are labeled as ``CN and joint overlap'' and ``ME and joint overlap'' for copy number and methylation respectively, in the supplementary tables 2-5. When the joint test leads to many extra discoveries, compared with the individual test, this proportion is small. If the proportion is close to 1, however, the joint test statistic mostly finds associations already identified by the individual test, suggesting that a single molecular profile drives effects on gene expression. 

The complement of this proportion (obtained by computing 1-proportion), the ratio of new discoveries with the joint test, points out clearly that large fractions of joint test new discoveries are yielded when compared with the ME test (figure 4).  This was expected, due to the stronger copy number effect that the methylation test does not capture on its own, but that the joint test does. Nevertheless, the advantage of the joint test is evident as there are new discoveries with the joint test for all chromosome arms.

The overlap ratios show that, as expected, a relatively large proportion of mRNA probes selected by joint tests is also selected by the CN test (80\% and 70\% for colon and breast, respectively - supplementary table 3). In contrast, selected joint tests are also selected with the ME test at considerably smaller proportions (29\% and 41\% for colon and breast, respectively - supplementary table 3). This can be again explained by the strong copy number effects and, as these are more pronounced in colon cancer than in breast cancer the joint test, compared to the CN test, leads to more extra discoveries in the breast cancer data than in the colon cancer data. Note that herewith when we refer to a  ``selected test'' we obviously mean a ``selected mRNA probe by the test''.


Results per chromosome arm mostly reflect the strong copy number effect, especially for colon cancer (figure 5, graphs on top row). 
Some special patterns appear: for 8p and 8q, mRNA probes selected by the joint test are, but for a fraction around 10\%, also selected by the CN test, for both cancers. The overlap between the ME and joint tests is smaller, between 20 and 27\% (supplementary tables 4 and 5). As most of the mRNA probes selected by the ME test are also selected by the CN test, especially for 8q, these arms display for both colon and breast cancers similar patterns to 20q and 13q for colon cancer (see supplementary figure 9 for an overview, and supplementary figure 11 for 8q). 


Another interesting case is 18p, where virtually all (99\%)  selected joint tests also yield significant CN tests for colon, whilst for breast cancer this proportion is 50\%. Furthermore, for colon cancer only 14\% of the selected joint tests are also selected  ME tests, whilst for breast cancer this proportion is 50\%. 
Indeed, in breast cancer CN and ME tests are selected for different mRNA probes: the overlap between significant CN and ME tests is 0 (supplementary figure 9). Thus, for genes on 18p, copy number changes drives gene expression regulation in colon cancer. In contrast, in breast cancer  methylation changes also drives gene expression regulation, independently of copy number.


Another interesting example is 19q, for which approximately 63 and 70\% of the selected joint tests were not selected by the CN test individually, for colon and breast cancer respectively. The same 19q displays the largest overlap (76 and 70\% for colon and breast, resp.) between selected joint and ME tests, leading us to conclude that methylation effects drive gene expression regulation on 19q, for both cancers.  Indeed,  methylation effects dominate the joint test statistic, although copy number effects are still found for genes that are not affected by methylation change (supplementary figure 12), with subsequently little overlap between selected CN and ME tests (supplementary figure 9).


That methylation plays a stronger role in gene expression regulation in breast cancer, compared with colon cancer, has thus been made clear. This is particularly noticeable when we examine the overlap proportions between selected joint and methylation tests; these represent strong enough effects, although they may in some cases overlap with selected copy number tests too. On chromosome 9, in particular, these proportions of overlap are strikingly larger in the breast cancer data (55 and 51\% for 9p and 9q, respectively), compared with the colon cancer data (23 and 21\% for 9p and 9q -- see supplementary tables 4 and 5).
Interestingly, one of the selected genes on 9p is {\sl CDKN2A}, which is mapped by six mRNA probes on the Agilent expression microarray used. Of these, two are  selected as associated with both the methylation and the joint tests in the breast cancer data, and two are just above the selection threshold (p-values $\leq$ 0.005) with these two tests. In contrast, in the colon cancer data only one probe  would be selected with a less stringent cut-off, by the methylation and the joint tests (p $\leq$  0.005). Indeed, methylation and gene expression are associated in both data sets, although the correlation is more widely and evenly spread across both methylation as mRNA-expression probes (supplementary figure 13).

In order to better understand the mechanisms regulating {\sl CDKN2A\/} expression, let us consider the mRNA probe \verb{A_23_P17356{ which has significant associations in breast, but not colon, cancer. A total of 49 breast cancer samples display DNA copy loss, whilst only 6 colon cancer samples display a small loss in this region (supplementary figures 14--15, top). Note that the mRNA probe is often not under-expressed for these samples. If we now sort samples in the methylation data according to for the copy number data heatmaps, we see that the samples displaying larger DNA copy loss (right-hand side) also display hypo-methylation of about half of the probes which, again, is only observed for breast cancer (supplementary figures 14--15, bottom). If we now look at scatterplots of \verb{A_23_P17356{ expression and the 19 methylation probes in the covariate set tests, we notice that many of those display negative association with mRNA expression in breast cancer, but not in colon, as expected of a functional hyper-methylation event affecting gene expression (supplementary figures 16--17). In particular, DNA loss is associated with hypo-methylation of some probes in breast cancer, but these probes are not correlated with mRNA expression. Indeed, methylation probes correlated with mRNA expression show no correlation with DNA copy number. So, in breast cancer methylation affects {\sl CDKN2A} expression, independently of DNA copy number change.

\subsubsection*{Tests with different conclusions}

Clearly mRNA probes for which the joint test statistic is selected (i.e., has $p\leq 0.001$) but one of the individual associations (CN or ME) is not, may mean that the other individual association is thus selected, driving the result of the joint test. This is indeed often the case, but not for all tests. Indeed, there is a small but non-negligible proportion of mRNA probes (0.6\% for colon and 1.2\% for breast) that is selected by the joint test, whilst that is not the case with either of the individual tests (see line ``Joint sel but not CN ME/joint sel" in supplementary tables 2--3). Such cases tend to reflect small effects that are spread across multiple covariates, in this case both copy number and methylation probes, and the collection of small effects leads to a result being selected by considering the larger covariate set. As such, these are worthy of further investigation. Looking at the results per chromosome arm, we find that 11p and 17p have relatively high proportions ( 4\% and 3\% respectively) of effects found only with the joint test statistic, for the breast cancer data (see supplementary figure 27 and supplementary table 5 in Additional File 2). 


There are of course mRNA probes that are selected only by an individual test, but not by the joint test. Indeed, around 21\% of the mRNA probes selected by CN tests are not selected by the joint test, for both colon and breast cancers. In such effect dilution occurs, the conclusion is obviously that copy number affects gene expression whilst methylation does not, to an extent that the copy number effect, either not very strong or only spanning a small subset of the copy number variables, is no longer enough to drive the result of the joint test statistic. From the mRNA probes selected by ME tests, between 15\% and 27\% are found not to be selected by the joint test, for colon and breast cancer respectively.


Most chromosome arms display similar, small dilution of copy number effects for colon and breast, such as for example 8q (figure 5, bottom-row graphs). This was partly expected due to the large overlap between joint and CN tests, in case of 8q almost total. There are exceptions, however: almost 100\% of the mRNA probes  on 18p selected by the joint test are also selected by the CN test in colon cancer, and yet 12\% of all selected mRNA probes with the CN test were diluted by the joint test statistic. For 19q, of all mRNA probes selected by the CN test, around 57\% for colon and 43\% for breast were not selected by the joint test statistic, whilst for the ME test these proportions were 1 and 8\% respectively. So, 19q is a chromosome arm where methylation effects dominate the joint test statistic and, as such, are diluted to a very small extent; on the other hand, about half of the individual copy number effects are diluted in the joint test statistic.

Overall, we observed more dilution of methylation effects in the joint test statistic with the breast cancer data, compared with colon (bottom-right hand-side graph in figure 5). This is likely due to the fact that the breast cancer displays more methylation effects but, as these effects are mild, they are more likely to be diluted in the joint test statistic. The largest methylation-effect dilutions in the breast cancer data were observed for 6q, 7p, 8p and 8q. We have already seen that, for 8p and 8q, at least 92\% of the mRNA probes selected by the joint test are also selected by the CN test, for both colon and breast cancer. This could be due to methylation having little effect on gene expression. What the dilution proportions tell us is that the copy number effect dominance comes at the cost of the methylation effect: between 18 and 31\% in colon cancer, and between 53 and 71\% in breast cancer, of the methylation effect is ``lost''. This is also evident from the gene-wise test graphs for these chromosome arms, as a set of p-values that is near zero for the ME test, but not for CN or joint tests (supplementary figures 11 and 19 for 8q and 6q, respectively).

\subsubsection*{Examples of effects found}

Our results highlight a variety of effects of copy number and methylation explaining gene expression variability. To illustrate this, we selected probes found with our tests and examined corresponding patterns in the data motivating the findings. For each test, we select probes for which a test had p-value $\leq 0.001$. 

Firstly, we looked for mRNA probes selected with the joint test, as well as both CN and ME tests, on 13q, 20q and 8q, in the colon cancer data. We further refined our search by requiring that the probe had CN test p-value $>0.1$ in the breast cancer data. This led to 56, 104 and 54 mRNA probes being selected, respectively on 13q, 20q and 8q. Secondly, we looked for probes that had both ME test and joint test p-values selected, 
on 19q in the colon cancer data. We again further refined the selection by selecting only probes that had ME test p-value $>0.1$ in the breast cancer data. This led to 98 mRNA expression probes. From each one of these four lists, we select a single probe to illustrate effects found, except for 13q and 20q from which we will examine two and four probes, respectively. Finally, we consider also one probe mapping to the gene {\sl CDKN2A\/}, for which p-values are just above cut-off for ME and joint tests, in colon cancer. A list of the probes selected, with annotation and tests results in both data sets, can be found in supplementary table 7.

The selection criteria used are likely to find mRNA probes with strong copy number effects on 13q, 20q and 8q, as we know these characterize the colon cancer data. Indeed, this is what we find, but with an interesting twist: copy number seems to explain  mRNA expression for only part of the samples, with in almost all cases a subset of samples with diploid copy number and yet varying mRNA expression. This is the case with genes {\sl GGT7, PIGU, NUFIP1,  WFDC2, SLC39A4, PCDH20\/} and {\sl  IFT52\/} (figure 6 and supplementary figures 20--25). Since these probes also were selected on the basis of the ME test, we expect that methylation is also regulating mRNA expression, in particular in diploid samples, in spite of the strong copy number effect. Indeed, we found that, not only samples with DNA copy gain also have less methylation, but also diploid samples more often display more methylation. For these samples, it seems that copy number and methylation have an additive effect on mRNA expression. Note, however, that in most cases, copy number and methylation do not explain mRNA expression completely, with a few samples displaying less methylation as well as a copy gain, and yet being under-expressed compared to the remaining samples. This suggests that a third mechanism is regulating mRNA expression for a subset of samples, for example down-regulating mRNA expression in samples with DNA copy gain and less than average methylation, or up-regulating mRNA expression in diploid samples with more than average methylation. In such cases, the event of up- or down-regulation of mRNA expression is thus achieved by different mechanisms depending on the sample, which illustrates the concept of complementary effects introduced in the simulation study.

It is interesting to note that the mRNA regulatory mechanism just described implies that DNA copy number and methylation are negatively correlated. This is the opposite from what we expect if a copy number change occurs at random, which would alter methylation in the same direction. This suggests again that the methylation changes are functional, since they neither can be a consequence of DNA copy gain nor are they likely to have occurred by chance (and not be functional) on the diploid samples.

Not all probes follow this pattern. Probe \verb{A_23_P17356{, mapping to gene {\sl GDAP1L1\/}, is interesting because copy number also separates samples into diploid and copy gain, but now copy gain samples are under-expressed compared with diploid ones (figure 7). Methylation still displays a (negative) association with gene expression, for at least two probes. It is possible that methylation is compensating for the DNA copy gain here, since 20q is very often gained in colon cancer.  

Finally, the probe selected on 19q on the basis of the ME and joint tests, mapping to the {\sl SLC7A9\/}, shows indeed methylation negatively correlated with mRNA expression (supplementary figure 26). As expected, DNA copy number is not correlated with mRNA expression. 

The probe chosen mapping to {\sl CDKN2A\/} displays a similar pattern to that for {\sl SLC7A9\/}, with a trend of over-expression for samples displaying more methylation by one or both methylation probes. On the other hand, almost all samples with under-expression  displayed more methylation with at least one of the two methylation probes  (supplementary figure 27). This was not observed for the same probe in breast cancer samples. We conclude that expression of {\sl CDKN2A\/} is often regulated via methylation, but that this can be best represented by different mRNA probes, for breast and colon cancer. Methylation probes showing strong association with mRNA expression may be the same, as is here the case. 

After observing that methylation plays an important role in gene-expression regulation in breast cancer, we also looked for genes that had, for the ME test, $p\leq 0.001$ and $p>0.1$ in breast and colon cancer, respectively. This yielded 439 genes. In order to make the selection even stricter, we required that each gene had at least two mRNA probes satisfying the criterion, which led to 274 genes. This list included gene {\sl MGC29506\/}, also known as {\sl MZB1\/}. It has been previously found to be frequently methylated in hepatocellular cancer \cite{matsumura2012}. Of the 4 mRNA probes mapping to it, two are methylated above average for all samples (data not shown). Another gene on this list is {\sl RAB40C\/}, a member of the RAS oncogene family. Of the 6 mRNA probes mapping to it, 2 are significant and exhibit a negative correlation between median copy number and mRNA expression (data not shown).

The above mechanisms illustrate effect types that can be found by our approach, which has squeezed relevant information from thousands of mRNA probes, copy number and methylation measurements and yielded a list of mRNAs for further investigation. Obviously, correlation does not mean causation: further research would be needed to verify which of the associations found indeed is functional. In particular, statistical significance need not lead to biological relevance: as with any statistical method, there are effects found of too small a magnitude to be biologically relevant.

\section*{Discussion}

We have proposed a test to find associations between a dependent variable and two or more sets of covariates. In the context of studies where multiple molecular profiles are available for each sample, as for example gene expression, copy number and methylation, such a test can be used per gene expression variable to test for associations with copy number and/or methylation. These help us better understand molecular mechanisms of gene expression regulation, individually and jointly, as we showed with the colon and breast cancer TCGA data. 

Other methods have been proposed to look for associations between one molecular profile, and two or more other profiles. One framework that has been used by various authors for the two-dataset case (one independent) and, by some, has been extended to the three-dataset case, is that of penalized canonical correlations \cite{waaijenborg2009},\cite{witten2009},\cite{lee2011}. This framework tries to find (linear) combinations of the variables in the molecular profiles datasets that are (most) strongly correlated. As such, they are of an exploratory nature, whereas our approach is meant for inference, pinpointing associations that are statistically significant. Indeed, canonical vectors essentially identify sets of variables that are most strongly correlated. While their correlation may be considered high, it could still be due to chance, especially considering the high-dimensionality and sparseness of the data sets. In contrast, our approach takes the sets of covariates as given, and tests the association with the dependent variable for statistical significance. 

In addition, due to their very nature sparse-canonical correlation-based approaches are computationally complex, and the complexity increases very quickly with each added independent dataset, as well illustrated by Lee \etal \cite{lee2011}.  Compounded with the high-dimensions of the datasets involved, this means that some of them do not easily scale up. In addition, it can be difficult to interpret the canonical vectors.

Bayesian approaches have also been proposed to analyse two datasets (one independent) together \cite{peng2010},\cite{richardson2011},\cite{khalili2011}, but as such approaches are naturally computationally complex, they often become prohibitively computationally complex when used on the whole genome, or for more than one independent data set. Richardson \etal\cite{richardson2011} try to address the complexity by proposing a more efficient algorithm for MCMC estimation, although they do not apply their method to more than two data sets in their paper. 

Vaske\cite{vaske2010} proposed another method to analyse multiple molecular profiles simultaneously. Their method consists first of finding associations between probes in different data sets and (sets of) pathways. Subsequently, their matrix of inferred associations between samples and pathways is used instead of the entire dataset in analyses. This leads to a dimension reduction, helping the later computational task of studying associations with clinical variables. Based on the reduced data set and on known possible interactions between molecules, directed graphs are built. This process involves discretization of the data set and of the interactions, to enable establishment of direction in the graph estimation. As such, their method relies on pathways used containing information relevant to the study, as well as biological information about molecular interactions within pathways. This makes for a much more structured method, that is of particular interest when specific pathways are under study. 

Our method is intentionally less structured. We look at all molecular data with an open mind, and are able to find regulatory mechanisms that may not be linked to any known pathway, or may affect genes involved in pathways only mildly.  So it is preferred in cases when one wishes to search for regulatory mechanisms affecting gene expression, or even molecular mechanisms associated as a set to a clinical variable. The lack of imposed structure makes it more flexible, since we can easily apply it to study the effect of DNA copy number and methylation together on expression levels of a gene, or a protein. Precisely the same model could have been used to study the effects of SNP genotypes and methylation too. Indeed, our approach could be used on the matrix of inferred pathway activities generated by the approach of Vaske \etal\cite{vaske2010}, instead of their proposed graphical model, where we would have ignored the interaction factors. Here all pathways could be used in a single set of covariates to study relationship with a clinical outcome, for example.

Our approach is computationally simple, involving at its most complicated matrix multiplications for the linear case. It is also flexible, in that it may correct for confounder effects. The extension from two sets of covariates to more sets is straightforward, once we ignore correlations between test statistics. Even if correlations were not to be ignored, a closed form for the test statistic is available once the number of covariate sets is fixed. 
One of the reasons why our approach is simpler is that the focus is on testing, rather than on model fitting, as do Bayesian approaches, or on finding canonical vectors. This allows for quick, simple and objective evaluation of results, and easy prioritization of found associations.

In the TCGA data examples, we tested for association between gene expression and  both copy number and methylation changes, in  colon and breast cancer samples. We observed  similarities and differences between these two cancer types. For some chromosome arms, such as 1q and 17q, proportions of discoveries are similar between the two cancers. This needs not mean that the same gene expression-regulating mechanisms are involved, but that genomic changes of the same type are involved at the same ratio. Also, copy number and methylation do not necessarily explain gene expression variation of the same genes, and the joint test finds additional associations compared to the individual tests. For other arms, such as 13q and 20q, there are clear differences between the genomic changes found to affect gene expression in colon and breast cancer. 

Some genes were selected to illustrate patterns found by the method, which showed some expression-regulating mechanisms driven by a combination of copy number and methylation changes. The gene selection aimed at illustrating differences found when chromosome arms results were compared between the two cancer types. Researchers interested in other selection criteria can apply those to the whole-genome results table provided  (Additional Files 3 and 4).

The pioneering articles of TCGA colon \cite{tcgacolon} and breast \cite{tcgabreast} cancer samples characterization have of course already given a detailed overview of molecular changes observed in these data sets. They have, however, not looked for complex regulatory mechanisms as we do here. So, our findings complement theirs, although we must be careful when extending the results from the samples subset used here to a larger set of cancer samples, as we cannot guarantee representativeness.

In the illustrations above we assumed that gene expression was the molecular phenotype of interest. Clearly that is not always the case, and other molecular phenotypes of interest include protein expression profiles. Our method can equally well be applied to such cases, with the reason for us not to have done so is the relatively limited number of samples with protein profiles available. In our experience, a minimum number of samples needed to yield reliable results is about 20 for the two-dataset case (one independent), but that of course is very much dependent on the signal-to-noise ratio in the data sets.

Our approach can also be used in a generalized-linear model context, where the dependent variables are related to the sets of covariates via a link function, such as the logarithm or the logit. This extension is straightforward, as we use the framework of the global test \cite{goeman2004},\cite{goeman2006}. This allows us for example to consider RNA-Seq data as dependent variables, if a suitable data transformation exists such that covariate effects can represent well variability  in the transformed RNA-Seq data. Ideally, we would like to represent the data variability using the negative binomial, as done for example by the BioConductor packages edgeR (\cite{robinson2007}) and DESeq (\cite{anders2010}), as it handles the data overdispersion (\cite{whitaker1914}).  Note, however, that this is not straightforward as the negative binomial distribution is not in the exponential family. This extension is beyond the objectives of this paper and will appear elsewhere. 

    Another application of interest is to study the impact of changes in multiple molecular profiles on patients' survival outcome, a variable formed by time from diagnosis to event and event information. This can be done by means of a Cox proportional-hazards model as an extension of the global test for this setting proposed earlier \cite{goeman2005}.

\section*{Conclusions}

We propose a method to find effects of multiple molecular profiles on a response, which can be a single clinical variable or a molecular profile. Our method is computationally simple, making it scalable for use over the entire genome. Its flexibility means it can be used with a variety of molecular types. It can be very useful to unravel complex gene expression-regulating mechanisms.


\begin{backmatter}

\section*{Competing interests}
  The authors declare that they have no competing interests.

\section*{Author's contributions}
    
  RXM first proposed the paper's concept, further developed the test for multiple covariate sets, performed the analyses and wrote the manuscript. 	LM has worked out the extension of the global test to two covariate sets. 
	JJG helped shape the discussion around the final analytical form of the test statistic, with insight in why ignoring the covariances seemed to be advantageous.
	JMB contributed to further development of the paper's concept in terms of biological applications, including designing the simulation study and interpreting results of the cancer data analysis.
	All authors read and approved of the final manuscript.

\section*{Acknowledgements}

  We are grateful for helpful comments and discussion from M. van de Wiel and W. van Wieringen. We also acknowledge the use of data produced by The Cancer Genome Atlas (TCGA) Research Network to illustrate methods introduced in this work.


\bibliographystyle{bmc-mathphys} 
  \bibliography{integration_model}      


\begin{thebibliography}{25}
\ifx \bisbn   \undefined \def \bisbn  #1{ISBN #1}\fi
\ifx \binits  \undefined \def \binits#1{#1}\fi
\ifx \bauthor  \undefined \def \bauthor#1{#1}\fi
\ifx \batitle  \undefined \def \batitle#1{#1}\fi
\ifx \bjtitle  \undefined \def \bjtitle#1{#1}\fi
\ifx \bvolume  \undefined \def \bvolume#1{\textbf{#1}}\fi
\ifx \byear  \undefined \def \byear#1{#1}\fi
\ifx \bissue  \undefined \def \bissue#1{#1}\fi
\ifx \bfpage  \undefined \def \bfpage#1{#1}\fi
\ifx \blpage  \undefined \def \blpage #1{#1}\fi
\ifx \burl  \undefined \def \burl#1{\textsf{#1}}\fi
\ifx \doiurl  \undefined \def \doiurl#1{\textsf{#1}}\fi
\ifx \betal  \undefined \def \betal{\textit{et al.}}\fi
\ifx \binstitute  \undefined \def \binstitute#1{#1}\fi
\ifx \binstitutionaled  \undefined \def \binstitutionaled#1{#1}\fi
\ifx \bctitle  \undefined \def \bctitle#1{#1}\fi
\ifx \beditor  \undefined \def \beditor#1{#1}\fi
\ifx \bpublisher  \undefined \def \bpublisher#1{#1}\fi
\ifx \bbtitle  \undefined \def \bbtitle#1{#1}\fi
\ifx \bedition  \undefined \def \bedition#1{#1}\fi
\ifx \bseriesno  \undefined \def \bseriesno#1{#1}\fi
\ifx \blocation  \undefined \def \blocation#1{#1}\fi
\ifx \bsertitle  \undefined \def \bsertitle#1{#1}\fi
\ifx \bsnm \undefined \def \bsnm#1{#1}\fi
\ifx \bsuffix \undefined \def \bsuffix#1{#1}\fi
\ifx \bparticle \undefined \def \bparticle#1{#1}\fi
\ifx \barticle \undefined \def \barticle#1{#1}\fi
\ifx \bconfdate \undefined \def \bconfdate #1{#1}\fi
\ifx \botherref \undefined \def \botherref #1{#1}\fi
\ifx \url \undefined \def \url#1{\textsf{#1}}\fi
\ifx \bchapter \undefined \def \bchapter#1{#1}\fi
\ifx \bbook \undefined \def \bbook#1{#1}\fi
\ifx \bcomment \undefined \def \bcomment#1{#1}\fi
\ifx \oauthor \undefined \def \oauthor#1{#1}\fi
\ifx \citeauthoryear \undefined \def \citeauthoryear#1{#1}\fi
\ifx \endbibitem  \undefined \def \endbibitem {}\fi
\ifx \bconflocation  \undefined \def \bconflocation#1{#1}\fi
\ifx \arxivurl  \undefined \def \arxivurl#1{\textsf{#1}}\fi
\csname PreBibitemsHook\endcsname

\bibitem{huang2012}
\begin{barticle}
\bauthor{\bsnm{Huang}, \binits{N.}},
\bauthor{\bsnm{Shah}, \binits{P.}},
\bauthor{\bsnm{Li}, \binits{C.}}:
\batitle{Lessons from a decade of integrating cancer copy number alterations
  with gene expression profiles}.
\bjtitle{Briefings in Bioinformatics}
\bvolume{13}(\bissue{3}),
\bfpage{305}--\blpage{316}
(\byear{2012}).
doi:\doiurl{10.1093/bib/bbr056}.
\bcomment{{PMID:} 21949216}
\end{barticle}
\endbibitem

\bibitem{menezes2009}
\begin{barticle}
\bauthor{\bsnm{Menezes}, \binits{R.}},
\bauthor{\bsnm{Boetzer}, \binits{M.}},
\bauthor{\bsnm{Sieswerda}, \binits{M.}},
\bauthor{\bparticle{van} \bsnm{Ommen}, \binits{G.}},
\bauthor{\bsnm{Boer}, \binits{J.}}:
\batitle{Integrated analysis of dna copy number and gene expression microarray
  data using gene sets}.
\bjtitle{BMC Bioinformatics}
\bvolume{10},
\bfpage{203}
(\byear{2009})
\end{barticle}
\endbibitem

\bibitem{stranger2007}
\begin{barticle}
\bauthor{\bsnm{Stranger}, \binits{B.}},
\bauthor{\bsnm{Forrest}, \binits{M.}},
\bauthor{\bsnm{Dunning}, \binits{M.}},
\bauthor{\bsnm{Ingle}, \binits{C.}},
\bauthor{\bsnm{Beazley}, \binits{C.}},
\bauthor{\bsnm{Thorne}, \binits{N.}},
\bauthor{\bsnm{Redon}, \binits{R.}},
\bauthor{\bsnm{Bird}, \binits{C.}},
\bauthor{\bparticle{de} \bsnm{Grassi}, \binits{A.}},
\bauthor{\bsnm{Lee}, \binits{C.}},
\bauthor{\bsnm{Tyler-Smith}, \binits{C.}},
\bauthor{\bsnm{Carter}, \binits{N.}},
\bauthor{\bsnm{Scherer}, \binits{S.}},
\bauthor{\bsnm{Tavar\'{e}}, \binits{S.}},
\bauthor{\bsnm{Deloukas}, \binits{P.}},
\bauthor{\bsnm{Hurles}, \binits{M.}},
\bauthor{\bsnm{Dermitzakis}, \binits{E.}}:
\batitle{Relative impact of nucleotide and copy number variation on gene
  expression phenotypes}.
\bjtitle{Science}
\bvolume{315},
\bfpage{848}--\blpage{853}
(\byear{2007})
\end{barticle}
\endbibitem

\bibitem{iterson2013}
\begin{botherref}
\oauthor{\bsnm{Iterson}, \binits{M.v.}},
\oauthor{\bsnm{Bervoets}, \binits{S.}},
\oauthor{\bsnm{Meijer}, \binits{E.d.}},
\oauthor{\bsnm{Buermans}, \binits{H.}},
\oauthor{\bsnm{Hoen}, \binits{P.}},
\oauthor{\bsnm{Menezes}, \binits{R.}},
\oauthor{\bsnm{Boer}, \binits{J.}}:
Integrated analysis of microrna and mrna expression: adding biological
  significance to microrna target predictions.
Nucleic Acids Research,
146
(2013).
{PMID:} 23771142
\end{botherref}
\endbibitem

\bibitem{waaijenborg2009}
\begin{barticle}
\bauthor{\bsnm{Waaijenborg}, \binits{S.}},
\bauthor{\bsnm{Zwinderman}, \binits{A.}}:
\batitle{Sparse canonical correlation analysis for identifying, connecting and
  completing gene-expression networks}.
\bjtitle{BMC Bioinformatics}
\bvolume{10},
\bfpage{315}
(\byear{2009})
\end{barticle}
\endbibitem

\bibitem{witten2009}
\begin{barticle}
\bauthor{\bsnm{Witten}, \binits{D.}},
\bauthor{\bsnm{Tibshirani}, \binits{R.}}:
\batitle{Extensions of sparse canonical correlation analysis with applications
  to genomic data}.
\bjtitle{Statistical Applications in Genetics and Molecular Biology}
\bvolume{8},
\bfpage{28}
(\byear{2009})
\end{barticle}
\endbibitem

\bibitem{goeman2004}
\begin{barticle}
\bauthor{\bsnm{Goeman}, \binits{J.}},
\bauthor{\bparticle{van~de} \bsnm{Geer}, \binits{S.}},
\bauthor{\bparticle{de} \bsnm{Kort}, \binits{F.}},
\bauthor{\bparticle{van} \bsnm{Houwelingen}, \binits{H.}}:
\batitle{A global test for groups of genes: testing association with a clinical
  outcome}.
\bjtitle{Bioinformatics}
\bvolume{20},
\bfpage{93}--\blpage{99}
(\byear{2004})
\end{barticle}
\endbibitem

\bibitem{lecessie1995}
\begin{barticle}
\bauthor{\bsnm{Le~Cessie}, \binits{S.}},
\bauthor{\bparticle{van} \bsnm{Houwelingen}, \binits{H.}}:
\batitle{Testing the fit of a regression-model via score tests in random
  effects models}.
\bjtitle{Biometrics}
\bvolume{51},
\bfpage{600}--\blpage{614}
(\byear{1995})
\end{barticle}
\endbibitem

\bibitem{goeman2005}
\begin{barticle}
\bauthor{\bsnm{Goeman}, \binits{J.}},
\bauthor{\bsnm{Oosting}, \binits{J.}},
\bauthor{\bsnm{Cleton-Jansen}, \binits{A.-M.}},
\bauthor{\bsnm{Anninga}, \binits{J.}},
\bauthor{\bparticle{van} \bsnm{Houwelingen}, \binits{H.}}:
\batitle{Testing association of a pathway with survival using gene expression
  data}.
\bjtitle{Bioinformatics}
\bvolume{21},
\bfpage{1950}--\blpage{1957}
(\byear{2005})
\end{barticle}
\endbibitem

\bibitem{goeman2006}
\begin{barticle}
\bauthor{\bsnm{Goeman}, \binits{J.}},
\bauthor{\bparticle{van~de} \bsnm{Geer}, \binits{S.}},
\bauthor{\bparticle{van} \bsnm{Houwelingen}, \binits{H.}}:
\batitle{Testing against a high dimensional alternative}.
\bjtitle{Journal of the Royal Statistical Society Series B}
\bvolume{68},
\bfpage{477}--\blpage{493}
(\byear{2006})
\end{barticle}
\endbibitem

\bibitem{goeman2011}
\begin{barticle}
\bauthor{\bsnm{Goeman}, \binits{J.}},
\bauthor{\bparticle{van} \bsnm{Houwelingen}, \binits{H.}},
\bauthor{\bsnm{Finos}, \binits{L.}}:
\batitle{Testing against a high dimensional alternative in the generalized
  linear model: asymptotic type i error control}.
\bjtitle{Biometrika}
\bvolume{98},
\bfpage{381}--\blpage{390}
(\byear{2011})
\end{barticle}
\endbibitem

\bibitem{refR2015}
\begin{bbook}
\bauthor{\bsnm{{R Core Team}}}:
\bbtitle{R: A Language and Environment for Statistical Computing}.
\bpublisher{R Foundation for Statistical Computing},
\blocation{Vienna, Austria}
(\byear{2015}).
\bcomment{R Foundation for Statistical Computing}.
\burl{https://www.R-project.org/}
\end{bbook}
\endbibitem

\bibitem{blackburn2015}
\begin{barticle}
\bauthor{\bsnm{Blackburn}, \binits{A.}},
\bauthor{\bsnm{Almeida}, \binits{M.}},
\bauthor{\bsnm{Dean}, \binits{A.}},
\bauthor{\bsnm{Curran}, \binits{J.}},
\bauthor{\bsnm{Johnson}, \binits{M.}},
\bauthor{\bsnm{Moses}, \binits{E.}},
\bauthor{\bsnm{Abraham}, \binits{L.}},
\bauthor{\bsnm{Carless}, \binits{M.}},
\bauthor{\bsnm{Dyer}, \binits{T.}},
\bauthor{\bsnm{Kumar}, \binits{S.}},
\bauthor{\bsnm{Almasy}, \binits{L.}},
\bauthor{\bsnm{Mahaney}, \binits{M.}},
\bauthor{\bsnm{Comuzzie}, \binits{A.}},
\bauthor{\bsnm{Williams-Blangero}, \binits{S.}},
\bauthor{\bsnm{Blangero}, \binits{J.}},
\bauthor{\bsnm{Lehman}, \binits{D.}},
\bauthor{\bsnm{Goring}, \binits{H.}}:
\batitle{Effects of copy number variable regions on local gene expression in
  white blood cells of mexican americans}.
\bjtitle{European Journal of Human Genetics}
\bvolume{23},
\bfpage{1229}--\blpage{1235}
(\byear{2015})
\end{barticle}
\endbibitem

\bibitem{bell2011}
\begin{barticle}
\bauthor{\bsnm{Bell}, \binits{J.}},
\bauthor{\bsnm{Pai}, \binits{A.}},
\bauthor{\bsnm{Pickrell}, \binits{J.}},
\bauthor{\bsnm{Gaffney}, \binits{D.}},
\bauthor{\bsnm{Pique-Regi}, \binits{R.}},
\bauthor{\bsnm{Degner}, \binits{J.}},
\bauthor{\bsnm{Gilad}, \binits{Y.}},
\bauthor{\bsnm{Pritchard}, \binits{J.}}:
\batitle{Dna methylation patterns associate with genetic and gene expression
  variation in hapmap cell lines}.
\bjtitle{Genome Biology}
\bvolume{12},
\bfpage{10}
(\byear{2011})
\end{barticle}
\endbibitem

\bibitem{matsumura2012}
\begin{barticle}
\bauthor{\bsnm{Matsumura}, \binits{S.}},
\bauthor{\bsnm{Imoto}, \binits{I.}},
\bauthor{\bsnm{Kozaki}, \binits{K.}},
\bauthor{\bsnm{Matsui}, \binits{T.}},
\bauthor{\bsnm{Muramatsu}, \binits{T.}},
\bauthor{\bsnm{Furuta}, \binits{M.}},
\bauthor{\bsnm{Tanaka}, \binits{S.}},
\bauthor{\bsnm{Sakamoto}, \binits{M.}},
\bauthor{\bsnm{Arii}, \binits{S.}},
\bauthor{\bsnm{Inazawa}, \binits{J.}}:
\batitle{Integrative array-based approach identifies mzb1 as a frequently
  methylated putative tumor suppressor in hepatocellular carcinoma}.
\bjtitle{Clinical Cancer Research}
\bvolume{18},
\bfpage{3541}--\blpage{3551}
(\byear{2012})
\end{barticle}
\endbibitem

\bibitem{lee2011}
\begin{barticle}
\bauthor{\bsnm{Le~Cessie}, \binits{S.}},
\bauthor{\bparticle{van} \bsnm{Houwelingen}, \binits{H.}}:
\batitle{Testing the fit of a regression-model via score tests in random
  effects models}.
\bjtitle{Statistical Applications in Genetics and Molecular Biology}
\bvolume{10},
\bfpage{30}
(\byear{2011})
\end{barticle}
\endbibitem

\bibitem{peng2010}
\begin{barticle}
\bauthor{\bsnm{Peng}, \binits{J.}},
\bauthor{\bsnm{Zhu}, \binits{J.}},
\bauthor{\bsnm{Bergamaschi}, \binits{A.}},
\bauthor{\bsnm{Han}, \binits{W.}},
\bauthor{\bsnm{Noh}, \binits{D.-Y.}},
\bauthor{\bsnm{Pollack}, \binits{J.}},
\bauthor{\bsnm{Wang}, \binits{P.}}:
\batitle{Regularized multivariate regression for identifying master predictors
  with application to integrative genomics study of breast cancer}.
\bjtitle{The Annals of Applied Statistics}
\bvolume{4}(\bissue{1}),
\bfpage{53}--\blpage{77}
(\byear{2010}).
doi:\doiurl{10.1214/09-AOAS271}
\end{barticle}
\endbibitem

\bibitem{richardson2011}
\begin{bchapter}
\bauthor{\bsnm{Richardson}, \binits{S.}},
\bauthor{\bsnm{Bottolo}, \binits{L.}},
\bauthor{\bsnm{Rosenthal}, \binits{J.}}:
\bctitle{Bayesian models for sparse regression analysis of high-dimensional
  data}.
In: \beditor{\bsnm{Bernardo}, \binits{J.}},
\beditor{\bsnm{Bayarri}, \binits{M.}},
\beditor{\bsnm{Berger}, \binits{J.}},
\beditor{\bsnm{Dawid}, \binits{A.}},
\beditor{\bsnm{Heckerman}, \binits{D.}},
\beditor{\bsnm{Smith}, \binits{A.}},
\beditor{\bsnm{West}, \binits{M.}} (eds.)
\bbtitle{Bayesian Statistics 9},
pp. \bfpage{397}--\blpage{420}.
\bpublisher{Oxford University Press}, \blocation{???}
(\byear{2011})
\end{bchapter}
\endbibitem

\bibitem{khalili2011}
\begin{barticle}
\bauthor{\bsnm{Khalili}, \binits{A.}},
\bauthor{\bsnm{Chen}, \binits{J.}},
\bauthor{\bsnm{Lin}, \binits{S.}}:
\batitle{Feature selection in finite mixture of sparse normal linear models in
  high-dimensional feature space}.
\bjtitle{Biostatistics}
\bvolume{12}(\bissue{1}),
\bfpage{156}--\blpage{172}
(\byear{2011}).
\bcomment{{PMID:} 20716532}
\end{barticle}
\endbibitem

\bibitem{vaske2010}
\begin{barticle}
\bauthor{\bsnm{Vaske}, \binits{C.}},
\bauthor{\bsnm{Benz}, \binits{S.}},
\bauthor{\bsnm{Sanborn}, \binits{J.}},
\bauthor{\bsnm{Earl}, \binits{D.}},
\bauthor{\bsnm{Szeto}, \binits{C.}},
\bauthor{\bsnm{Zhu}, \binits{J.}},
\bauthor{\bsnm{Haussler}, \binits{D.}},
\bauthor{\bsnm{Stuart}, \binits{J.}}:
\batitle{Inference of patient-specific pathway activities from
  multi-dimensional cancer genomics data using {PARADIGM}}.
\bjtitle{Bioinformatics}
\bvolume{26}(\bissue{12}),
\bfpage{237}--\blpage{245}
(\byear{2010}).
\bcomment{{PMID:} 20529912}
\end{barticle}
\endbibitem

\bibitem{tcgacolon}
\begin{barticle}
\bauthor{\bsnm{Network}, \binits{C.G.A.}}:
\batitle{Comprehensive molecular characterization of human colon and rectal
  cancer}.
\bjtitle{Nature}
\bvolume{487},
\bfpage{330}--\blpage{337}
(\byear{2012})
\end{barticle}
\endbibitem

\bibitem{tcgabreast}
\begin{barticle}
\bauthor{\bsnm{Network}, \binits{C.G.A.}}:
\batitle{Comprehensive molecular portraits of human breast tumours}.
\bjtitle{Nature}
\bvolume{490},
\bfpage{61}--\blpage{70}
(\byear{2012})
\end{barticle}
\endbibitem

\bibitem{robinson2007}
\begin{barticle}
\bauthor{\bsnm{Robinson}, \binits{M.}},
\bauthor{\bsnm{Smyth}, \binits{G.}}:
\batitle{Moderated statistical tests for assessing differences in tag
  abundance}.
\bjtitle{Bioinformatics}
\bvolume{23}(\bissue{21}),
\bfpage{2881}--\blpage{2887}
(\byear{2007}).
doi:\doiurl{10.1093/bioinformatics/btm453}.
\bcomment{{PMID:} 17881408}.
Accessed 2013-12-05
\end{barticle}
\endbibitem

\bibitem{anders2010}
\begin{barticle}
\bauthor{\bsnm{Anders}, \binits{S.}},
\bauthor{\bsnm{Huber}, \binits{W.}}:
\batitle{Differential expression analysis for sequence count data}.
\bjtitle{Genome Biology}
\bvolume{11}(\bissue{10}),
\bfpage{106}
(\byear{2010})
\end{barticle}
\endbibitem

\bibitem{whitaker1914}
\begin{barticle}
\bauthor{\bsnm{Whitaker}, \binits{L.}}:
\batitle{On the poisson law of small numbers}.
\bjtitle{Biometrika}
\bvolume{10},
\bfpage{36}--\blpage{71}
(\byear{1914})
\end{barticle}
\endbibitem

\end{thebibliography}

\newcommand{\BMCxmlcomment}[1]{}

\BMCxmlcomment{

<refgrp>

<bibl id="B1">
  <title><p>Lessons from a decade of integrating cancer copy number alterations
  with gene expression profiles</p></title>
  <aug>
    <au><snm>Huang</snm><fnm>N</fnm></au>
    <au><snm>Shah</snm><fnm>PK</fnm></au>
    <au><snm>Li</snm><fnm>C</fnm></au>
  </aug>
  <source>Briefings in Bioinformatics</source>
  <pubdate>2012</pubdate>
  <volume>13</volume>
  <issue>3</issue>
  <fpage>305</fpage>
  <lpage>-316</lpage>
  <note>{PMID:} 21949216</note>
</bibl>

<bibl id="B2">
  <title><p>Integrated analysis of DNA copy number and gene expression
  microarray data using gene sets</p></title>
  <aug>
    <au><snm>Menezes</snm><fnm>RX</fnm></au>
    <au><snm>Boetzer</snm><fnm>M</fnm></au>
    <au><snm>Sieswerda</snm><fnm>M</fnm></au>
    <au><snm>Ommen</snm><fnm>G</fnm></au>
    <au><snm>Boer</snm><fnm>JM</fnm></au>
  </aug>
  <source>BMC Bioinformatics</source>
  <pubdate>2009</pubdate>
  <volume>10</volume>
  <fpage>203</fpage>
</bibl>

<bibl id="B3">
  <title><p>Relative impact of nucleotide and copy number variation on gene
  expression phenotypes</p></title>
  <aug>
    <au><snm>Stranger</snm><fnm>BE</fnm></au>
    <au><snm>Forrest</snm><fnm>MS</fnm></au>
    <au><snm>Dunning</snm><fnm>M</fnm></au>
    <au><snm>Ingle</snm><fnm>CE</fnm></au>
    <au><snm>Beazley</snm><fnm>C</fnm></au>
    <au><snm>Thorne</snm><fnm>N</fnm></au>
    <au><snm>Redon</snm><fnm>R</fnm></au>
    <au><snm>Bird</snm><fnm>CP</fnm></au>
    <au><snm>Grassi</snm><fnm>A</fnm></au>
    <au><snm>Lee</snm><fnm>C</fnm></au>
    <au><snm>Tyler Smith</snm><fnm>C</fnm></au>
    <au><snm>Carter</snm><fnm>N</fnm></au>
    <au><snm>Scherer</snm><fnm>SW</fnm></au>
    <au><snm>Tavar\'{e}</snm><fnm>S</fnm></au>
    <au><snm>Deloukas</snm><fnm>P</fnm></au>
    <au><snm>Hurles</snm><fnm>ME</fnm></au>
    <au><snm>Dermitzakis</snm><fnm>ET</fnm></au>
  </aug>
  <source>Science</source>
  <pubdate>2007</pubdate>
  <volume>315</volume>
  <fpage>848</fpage>
  <lpage>-853</lpage>
</bibl>

<bibl id="B4">
  <title><p>Integrated analysis of microRNA and mRNA expression: adding
  biological significance to microRNA target predictions</p></title>
  <aug>
    <au><snm>Iterson</snm><fnm>Mv</fnm></au>
    <au><snm>Bervoets</snm><fnm>S</fnm></au>
    <au><snm>Meijer</snm><fnm>Ed</fnm></au>
    <au><snm>Buermans</snm><fnm>HP</fnm></au>
    <au><snm>Hoen</snm><fnm>PAC</fnm></au>
    <au><snm>Menezes</snm><fnm>RX</fnm></au>
    <au><snm>Boer</snm><fnm>JM</fnm></au>
  </aug>
  <source>Nucleic Acids Research</source>
  <pubdate>2013</pubdate>
  <fpage>e146</fpage>
  <url>http://nar.oxfordjournals.org/content/early/2013/06/14/nar.gkt525</url>
  <note>{PMID:} 23771142</note>
</bibl>

<bibl id="B5">
  <title><p>Sparse canonical correlation analysis for identifying, connecting
  and completing gene-expression networks</p></title>
  <aug>
    <au><snm>Waaijenborg</snm><fnm>S</fnm></au>
    <au><snm>Zwinderman</snm><fnm>AH</fnm></au>
  </aug>
  <source>BMC Bioinformatics</source>
  <pubdate>2009</pubdate>
  <volume>10</volume>
  <fpage>315</fpage>
</bibl>

<bibl id="B6">
  <title><p>Extensions of sparse canonical correlation analysis with
  applications to genomic data</p></title>
  <aug>
    <au><snm>Witten</snm><fnm>DM</fnm></au>
    <au><snm>Tibshirani</snm><fnm>RJ</fnm></au>
  </aug>
  <source>Statistical Applications in Genetics and Molecular Biology</source>
  <pubdate>2009</pubdate>
  <volume>8</volume>
  <fpage>28</fpage>
</bibl>

<bibl id="B7">
  <title><p>A global test for groups of genes: testing association with a
  clinical outcome</p></title>
  <aug>
    <au><snm>Goeman</snm><fnm>JJ</fnm></au>
    <au><snm>Geer</snm><fnm>SA</fnm></au>
    <au><snm>Kort</snm><fnm>F</fnm></au>
    <au><snm>Houwelingen</snm><fnm>HC</fnm></au>
  </aug>
  <source>Bioinformatics</source>
  <pubdate>2004</pubdate>
  <volume>20</volume>
  <fpage>93</fpage>
  <lpage>-99</lpage>
</bibl>

<bibl id="B8">
  <title><p>Testing the Fit of A Regression-Model Via Score Tests in Random
  Effects Models</p></title>
  <aug>
    <au><snm>Le Cessie</snm><fnm>S</fnm></au>
    <au><snm>Houwelingen</snm><fnm>HC</fnm></au>
  </aug>
  <source>Biometrics</source>
  <pubdate>1995</pubdate>
  <volume>51</volume>
  <fpage>600</fpage>
  <lpage>-614</lpage>
</bibl>

<bibl id="B9">
  <title><p>Testing association of a pathway with survival using gene
  expression data</p></title>
  <aug>
    <au><snm>Goeman</snm><fnm>JJ</fnm></au>
    <au><snm>Oosting</snm><fnm>J</fnm></au>
    <au><snm>Cleton Jansen</snm><fnm>A M</fnm></au>
    <au><snm>Anninga</snm><fnm>JK</fnm></au>
    <au><snm>Houwelingen</snm><fnm>HC</fnm></au>
  </aug>
  <source>Bioinformatics</source>
  <pubdate>2005</pubdate>
  <volume>21</volume>
  <fpage>1950</fpage>
  <lpage>-1957</lpage>
</bibl>

<bibl id="B10">
  <title><p>Testing against a high dimensional alternative</p></title>
  <aug>
    <au><snm>Goeman</snm><fnm>JJ</fnm></au>
    <au><snm>Geer</snm><fnm>SA</fnm></au>
    <au><snm>Houwelingen</snm><fnm>HC</fnm></au>
  </aug>
  <source>Journal of the Royal Statistical Society Series B</source>
  <pubdate>2006</pubdate>
  <volume>68</volume>
  <fpage>477</fpage>
  <lpage>-493</lpage>
</bibl>

<bibl id="B11">
  <title><p>Testing against a high dimensional alternative in the generalized
  linear model: asymptotic type I error control</p></title>
  <aug>
    <au><snm>Goeman</snm><fnm>JJ</fnm></au>
    <au><snm>Houwelingen</snm><fnm>HC</fnm></au>
    <au><snm>Finos</snm><fnm>L</fnm></au>
  </aug>
  <source>Biometrika</source>
  <pubdate>2011</pubdate>
  <volume>98</volume>
  <fpage>381</fpage>
  <lpage>-390</lpage>
</bibl>

<bibl id="B12">
  <title><p>R: A Language and Environment for Statistical Computing</p></title>
  <aug>
    <au><cnm>{R Core Team}</cnm></au>
  </aug>
  <publisher>Vienna, Austria</publisher>
  <pubdate>2015</pubdate>
  <url>https://www.R-project.org/</url>
</bibl>

<bibl id="B13">
  <title><p>Effects of copy number variable regions on local gene expression in
  white blood cells of Mexican Americans</p></title>
  <aug>
    <au><snm>Blackburn</snm><fnm>A</fnm></au>
    <au><snm>Almeida</snm><fnm>M</fnm></au>
    <au><snm>Dean</snm><fnm>A</fnm></au>
    <au><snm>Curran</snm><fnm>JE</fnm></au>
    <au><snm>Johnson</snm><fnm>MP</fnm></au>
    <au><snm>Moses</snm><fnm>EK</fnm></au>
    <au><snm>Abraham</snm><fnm>LJ</fnm></au>
    <au><snm>Carless</snm><fnm>MA</fnm></au>
    <au><snm>Dyer</snm><fnm>TD</fnm></au>
    <au><snm>Kumar</snm><fnm>S</fnm></au>
    <au><snm>Almasy</snm><fnm>L</fnm></au>
    <au><snm>Mahaney</snm><fnm>MC</fnm></au>
    <au><snm>Comuzzie</snm><fnm>A</fnm></au>
    <au><snm>Williams Blangero</snm><fnm>S</fnm></au>
    <au><snm>Blangero</snm><fnm>J</fnm></au>
    <au><snm>Lehman</snm><fnm>DM</fnm></au>
    <au><snm>Goring</snm><fnm>HHH</fnm></au>
  </aug>
  <source>European Journal of Human Genetics</source>
  <pubdate>2015</pubdate>
  <volume>23</volume>
  <fpage>1229</fpage>
  <lpage>1235</lpage>
</bibl>

<bibl id="B14">
  <title><p>DNA methylation patterns associate with genetic and gene expression
  variation in HapMap cell lines</p></title>
  <aug>
    <au><snm>Bell</snm><fnm>JT</fnm></au>
    <au><snm>Pai</snm><fnm>AA</fnm></au>
    <au><snm>Pickrell</snm><fnm>JK</fnm></au>
    <au><snm>Gaffney</snm><fnm>DJ</fnm></au>
    <au><snm>Pique Regi</snm><fnm>R</fnm></au>
    <au><snm>Degner</snm><fnm>JF</fnm></au>
    <au><snm>Gilad</snm><fnm>Y</fnm></au>
    <au><snm>Pritchard</snm><fnm>JK</fnm></au>
  </aug>
  <source>Genome Biology</source>
  <pubdate>2011</pubdate>
  <volume>12</volume>
  <fpage>R10</fpage>
</bibl>

<bibl id="B15">
  <title><p>Integrative Array-Based Approach Identifies MZB1 as a Frequently
  Methylated Putative Tumor Suppressor in Hepatocellular Carcinoma</p></title>
  <aug>
    <au><snm>Matsumura</snm><fnm>S</fnm></au>
    <au><snm>Imoto</snm><fnm>I</fnm></au>
    <au><snm>Kozaki</snm><fnm>K</fnm></au>
    <au><snm>Matsui</snm><fnm>T</fnm></au>
    <au><snm>Muramatsu</snm><fnm>T</fnm></au>
    <au><snm>Furuta</snm><fnm>M</fnm></au>
    <au><snm>Tanaka</snm><fnm>S</fnm></au>
    <au><snm>Sakamoto</snm><fnm>M</fnm></au>
    <au><snm>Arii</snm><fnm>S</fnm></au>
    <au><snm>Inazawa</snm><fnm>J</fnm></au>
  </aug>
  <source>Clinical Cancer Research</source>
  <pubdate>2012</pubdate>
  <volume>18</volume>
  <fpage>3541</fpage>
  <lpage>-3551</lpage>
</bibl>

<bibl id="B16">
  <title><p>Testing the Fit of A Regression-Model Via Score Tests in Random
  Effects Models</p></title>
  <aug>
    <au><snm>Le Cessie</snm><fnm>S</fnm></au>
    <au><snm>Houwelingen</snm><fnm>HC</fnm></au>
  </aug>
  <source>Statistical Applications in Genetics and Molecular Biology</source>
  <pubdate>2011</pubdate>
  <volume>10</volume>
  <fpage>30</fpage>
</bibl>

<bibl id="B17">
  <title><p>Regularized multivariate regression for identifying master
  predictors with application to integrative genomics study of breast
  cancer</p></title>
  <aug>
    <au><snm>Peng</snm><fnm>J</fnm></au>
    <au><snm>Zhu</snm><fnm>J</fnm></au>
    <au><snm>Bergamaschi</snm><fnm>A</fnm></au>
    <au><snm>Han</snm><fnm>W</fnm></au>
    <au><snm>Noh</snm><fnm>D Y</fnm></au>
    <au><snm>Pollack</snm><fnm>JR</fnm></au>
    <au><snm>Wang</snm><fnm>P</fnm></au>
  </aug>
  <source>The Annals of Applied Statistics</source>
  <pubdate>2010</pubdate>
  <volume>4</volume>
  <issue>1</issue>
  <fpage>53</fpage>
  <lpage>-77</lpage>
</bibl>

<bibl id="B18">
  <title><p>Bayesian models for sparse regression analysis of high-dimensional
  data</p></title>
  <aug>
    <au><snm>Richardson</snm><fnm>S</fnm></au>
    <au><snm>Bottolo</snm><fnm>L</fnm></au>
    <au><snm>Rosenthal</snm><fnm>JS</fnm></au>
  </aug>
  <source>Bayesian Statistics 9</source>
  <publisher>Oxford University Press</publisher>
  <editor>Bernardo, JM and Bayarri, MJ and Berger, JO and Dawid, AP and
  Heckerman, D and Smith, AFM and West, M</editor>
  <pubdate>2011</pubdate>
  <fpage>397</fpage>
  <lpage>-420</lpage>
</bibl>

<bibl id="B19">
  <title><p>Feature selection in finite mixture of sparse normal linear models
  in high-dimensional feature space</p></title>
  <aug>
    <au><snm>Khalili</snm><fnm>A</fnm></au>
    <au><snm>Chen</snm><fnm>J</fnm></au>
    <au><snm>Lin</snm><fnm>S</fnm></au>
  </aug>
  <source>Biostatistics</source>
  <pubdate>2011</pubdate>
  <volume>12</volume>
  <issue>1</issue>
  <fpage>156</fpage>
  <lpage>-172</lpage>
  <note>{PMID:} 20716532</note>
</bibl>

<bibl id="B20">
  <title><p>Inference of patient-specific pathway activities from
  multi-dimensional cancer genomics data using {PARADIGM}</p></title>
  <aug>
    <au><snm>Vaske</snm><fnm>CJ</fnm></au>
    <au><snm>Benz</snm><fnm>SC</fnm></au>
    <au><snm>Sanborn</snm><fnm>JZ</fnm></au>
    <au><snm>Earl</snm><fnm>D</fnm></au>
    <au><snm>Szeto</snm><fnm>C</fnm></au>
    <au><snm>Zhu</snm><fnm>J</fnm></au>
    <au><snm>Haussler</snm><fnm>D</fnm></au>
    <au><snm>Stuart</snm><fnm>JM</fnm></au>
  </aug>
  <source>Bioinformatics</source>
  <pubdate>2010</pubdate>
  <volume>26</volume>
  <issue>12</issue>
  <fpage>i237</fpage>
  <lpage>-i245</lpage>
  <note>{PMID:} 20529912</note>
</bibl>

<bibl id="B21">
  <title><p>Comprehensive molecular characterization of human colon and rectal
  cancer</p></title>
  <aug>
    <au><snm>Network</snm><fnm>CGA</fnm></au>
  </aug>
  <source>Nature</source>
  <pubdate>2012</pubdate>
  <volume>487</volume>
  <fpage>330</fpage>
  <lpage>-337</lpage>
</bibl>

<bibl id="B22">
  <title><p>Comprehensive molecular portraits of human breast
  tumours</p></title>
  <aug>
    <au><snm>Network</snm><fnm>CGA</fnm></au>
  </aug>
  <source>Nature</source>
  <pubdate>2012</pubdate>
  <volume>490</volume>
  <fpage>61</fpage>
  <lpage>-70</lpage>
</bibl>

<bibl id="B23">
  <title><p>Moderated statistical tests for assessing differences in tag
  abundance</p></title>
  <aug>
    <au><snm>Robinson</snm><fnm>MD</fnm></au>
    <au><snm>Smyth</snm><fnm>GK</fnm></au>
  </aug>
  <source>Bioinformatics</source>
  <pubdate>2007</pubdate>
  <volume>23</volume>
  <issue>21</issue>
  <fpage>2881</fpage>
  <lpage>-2887</lpage>
  <url>http://bioinformatics.oxfordjournals.org/content/23/21/2881</url>
  <note>{PMID:} 17881408</note>
</bibl>

<bibl id="B24">
  <title><p>Differential expression analysis for sequence count
  data</p></title>
  <aug>
    <au><snm>Anders</snm><fnm>S</fnm></au>
    <au><snm>Huber</snm><fnm>W</fnm></au>
  </aug>
  <source>Genome Biology</source>
  <pubdate>2010</pubdate>
  <volume>11</volume>
  <issue>10</issue>
  <fpage>R106</fpage>
</bibl>

<bibl id="B25">
  <title><p>On the Poisson law of small numbers</p></title>
  <aug>
    <au><snm>Whitaker</snm><fnm>L</fnm></au>
  </aug>
  <source>Biometrika</source>
  <pubdate>1914</pubdate>
  <volume>10</volume>
  <fpage>36</fpage>
  <lpage>-71</lpage>
</bibl>

</refgrp>
} 



\section*{Figures}
  \begin{figure}[h!]
  \caption{\csentence{ROC curves, simulation study 1.}
      ROC curves to evaluate power to find four different types of effects between one dependent variable and two independent covariates sets, $N=100$, Simulation Study 1. Purple lines represent curves for the the two-sets test statistic including the correlation between individual test statistics. Pink lines represent curves for the test statistic computed ignoring the correlation.}
      \end{figure}


\begin{figure}[h!]
  \caption{\csentence{ROC curves, simulation study 2.}
      ROC curves obtained with the test statistics including correlations between single-set test statistic $T^2_{XZ}$, and with the simplified test statistic $T^2_X+T^2_Z$ that ignore these correlations ($N=100$), for the Simulation Study 2.}
      \end{figure}


\begin{figure}[h!]
  \caption{\csentence{Comparison of copy number and methylation effects.}
      Proportion of genes selected by their expression association with copy number (left) and with methylation (right), for colon (x-axis) and breast (y-axis) cancers, per chromosome arm.}
      \end{figure}


\begin{figure}[h!]
  \caption{\csentence{Joint test new discoveries.}
       Computed as the proportion of genes with selected expression association with the joint set of covariates, that was found not to be selected by  either copy number (left-hand side, blue shades) or methylation (right-hand side, purple shades) separately. Top: colon cancer. Bottom: breast cancer.}
      \end{figure}


\begin{figure}[h!]
  \caption{\csentence{Comparing individual and joint tests: overlap and dilution.}
      Top row: overlap between individual and joint tests. Computed as the proportion of genes selected by both its expression being associated with the joint set of covariates, formed by both copy number and expression, as well as by each single set of covariates, per chromosome arm. This proportion is relative to all selected joint tests.  Bottom row: dilution. Computed as the proportion of selected genes for expression association with one set of covariates (either copy number or methylation), that are not selected with the joint set of covariates, formed by both copy number and expression, per chromosome arm. The proportion is relative to all selected tests for association with either copy number (bottom-left) or methylation (bottom-right).}
      \end{figure}


\begin{figure}[h!]
  \caption{\csentence{Gene {\sl GGT7\/}: expression, methylation and copy number.}
      mRNA expression (y-axis) of one probe mapping to gene {\sl GGT7\/} in the colon cancer data. Top graphs: methylation values (logit of beta values) for chosen probes within 50Kb of the gene start site; point colours represent the dychotomized copy number (see legend and dashed line in the graph right below), with blue corresponding to approximately diploid number of copies; the vertical dashed line represents the cut-off used to separate samples with more or less methylation. Middle graphs: median copy number values across all measurements within the 1 Mb region around the gene start site; point colours represent the dychotomized methylation (see legend and dashed line in the graph right above), with blue corresponding to less methylation; the vertical dashed line represents the cut-off used to separate samples approximately diploid from those. Bottom graph: mRNA expression for all samples, sorted from the smallest to the largest; three plotting symbols are used per sample to convey dychotomized copy number (star), methylation probe 1 (upwards triangle) and methylation probe 2 (downwards triangle); symbol colours are blue for lower values (diploid copy number, less methylation) or pink (copy gain, more methylation), as before.}
      \end{figure}


\begin{figure}[h!]
  \caption{\csentence{Gene {\sl GDAP1L1\/}: expression, methylation and copy number.}
      mRNA expression (y-axis) of one probe mapping to gene {\sl GDAP1L1\/} in the colon cancer data. Top graphs: methylation values (logit of beta values) for chosen probes within 50Kb of the gene start site; point colours represent the dychotomized copy number (see legend and dashed line in the graph right below), with blue corresponding to approximately diploid number of copies; the vertical dashed line represents the cut-off used to separate samples with more or less methylation. Middle graphs: median copy number values across all measurements within the 1 Mb region around the gene start site; point colours represent the dychotomized methylation (see legend and dashed line in the graph right above), with blue corresponding to less methylation; the vertical dashed line represents the cut-off used to separate samples approximately diploid from those. Bottom graph: mRNA expression for all samples, sorted from the smallest to the largest; three plotting symbols are used per sample to convey dychotomized copy number (star), methylation probe 1 (upwards triangle) and methylation probe 2 (downwards triangle); symbol colours are blue for lower values (diploid copy number, less methylation) or pink (copy gain, more methylation), as before.}
      \end{figure}



\section*{Additional Files}
  \subsection*{Additional file 1 --- Details in development of the test statistic} 
    This .pdf file includes details on how the test statistic is obtained, including a general functional form for the test statistic for any number $M$ of covariates set and the expression obtained for $M=3$. 
    

  \subsection*{Additional file 2 --- Supplementary tables and figures}
  
  This .pdf file contains details about the simulation study setup, as well as all supplementary figures and tables.

\subsection*{Additional file 3 -- Table of all results for the colon cancer data set}


\subsection*{Additional file 4 -- Table of all results for the breast cancer data set}


\end{backmatter}
\end{document}